\documentclass[twoside]{article}
\usepackage{qic}
\usepackage{graphicx}
\usepackage{amsmath}
\usepackage{amssymb}
\usepackage{bbm}
\usepackage{needspace}
\usepackage{cite}

\newcommand{\Tr}{\mathrm{Tr}}
\newcommand{\norm}[1]{\left\| #1 \right\|}
\newcommand{\ie}{\emph{i.e.} }
\newcommand{\rank}{\mathrm{rank}}
\newcommand{\bra}[1]{\langle #1 |}
\newcommand{\ket}[1]{| #1 \rangle}
\newcommand{\ball}[1]{\mathcal{B}^{\epsilon} (#1)}
\newcommand{\supp}{\mathrm{supp}}

\begin{document}
\setlength{\textheight}{8.0truein}    %FOR 2ND PAGE ONWARDS

\runninghead{An intuitive proof of the data processing inequality}
            {Normand J. Beaudry and Renato Renner}

\normalsize\textlineskip
\thispagestyle{empty}
\setcounter{page}{1}

%\copyrightheading{Vol.}{No.}{Year}{Page Nos.}
%\copyrightheading{0}{0}{2011}{000--000}

%\vspace*{0.88truein}

\alphfootnote

\fpage{1}

\centerline{\bf
AN INTUITIVE PROOF OF THE DATA PROCESSING INEQUALITY}
\vspace*{0.37truein}
\centerline{\footnotesize
NORMAND J. BEAUDRY AND RENATO RENNER}
\vspace*{0.015truein}
\centerline{\footnotesize\it Institute for Theoretical Physics, ETH Zurich, Wolfgang-Pauli-Str. 27}
\baselineskip=10pt
\centerline{\footnotesize\it 8093 Zurich, Switzerland}
%\vspace*{10pt}
%\centerline{\footnotesize 
%RENATO RENNER}
%\vspace*{0.015truein}
%\centerline{\footnotesize\it Institute for Theoretical Physics, ETH Zurich, Wolfgang-Pauli-Str. 27}
%\baselineskip=10pt
%\centerline{\footnotesize\it 8093 Zurich, Switzerland}
%\vspace*{0.225truein}
%\publisher{July 21, 2011}{(revised date)}

\vspace*{0.21truein}

\abstracts{
The data processing inequality (DPI) is a fundamental feature of information theory. Informally it states that you cannot increase the information content of a quantum system by acting on it with a local physical operation. When the smooth min-entropy is used as the relevant information measure, then the DPI follows immediately from the definition of the entropy. The DPI for the von Neumann entropy is then obtained by specializing the DPI for the smooth min-entropy by using the quantum asymptotic equipartition property (QAEP). We provide a short proof of the QAEP and therefore obtain a self-contained proof of the DPI for the von Neumann entropy.
}{}{}

\vspace*{10pt}

\keywords{Entropy, Information Theory}
\vspace*{3pt}
%\communicate{to be filled by the Editorial}

\vspace*{1pt}\textlineskip
\section{Introduction}
\noindent
The data processing inequality (DPI) has an intuitive interpretation: the information content in a quantum system cannot increase by performing local data processing on that system. It is an extremely useful property that is used extensively in quantum information \cite{nielsen00}. The DPI is known to hold for different entropy measures, and is stated generally as
\begin{equation}\label{eq:DPI}
\bar{H}(A|BC)_{\rho} \leq \bar{H}(A|B)_{\rho},
\end{equation}
where $\bar{H}(A|B)_{\rho}$ is a conditional entropic information measure of the state $\rho_{AB}$. Conditional entropy measures characterize the uncertainty about a system $A$ given a system $B$. The DPI is typically stated for the case where the local operation is a partial trace (i.e. a joint system $(B,C)$ is reduced to the system $B$), but this can be generalized to any physical operation.\footnote{The Stinespring dilation allows for any completely positive trace preserving (CPTP) map to be decomposed into a unitary followed by a partial trace. Since entropy measures are generally invariant under unitaries, the DPI applies to any CPTP map applied to the system $BC$.}

In particular, the DPI holds for one of the most widely used entropy measures: the conditional von Neumann entropy, $H(A|B)_{\rho}$ \cite{vonneumann55}. It is defined for normalized density operators acting on a bipartite Hilbert space $\mathcal{H}_{AB}$, $\rho \in S_{=}(\mathcal{H}_{AB})$ (where $S_{=}(\mathcal{H}):=\{\rho\in \mathcal{P}(\mathcal{H})|,\Tr(\rho)=1\}$ and $\mathcal{P}(\mathcal{H})$ is the set of positive semi-definite operators on $\mathcal{H}$), as $H(A|B)_{\rho} := H(AB)_{\rho} - H(B)_{\rho}$, where $H(A)_{\rho} := - \Tr(\rho_{A} \log \rho_{A})$ (all logarithms are taken to the base 2). For simplicity, we will not place the labels on density operators to denote which space they act on when it is clear from the context. Also, Eq.~\ref{eq:DPI} for the von Neumann entropy is equivalent to its strong subadditivity: $H(ABC)_{\rho}+H(B)_{\rho}\leq H(AB)_{\rho}+H(BC)_{\rho}$.

The first proofs of the DPI for the von Neumann entropy relied on abstract operator properties \cite{lieb73,lieb73b,simon79}. Recently these proofs have been simplified \cite{nielsen05,petz86,ruskai07}. Other approaches have used the operational meaning of the von Neumann entropy \cite{horodecki06,horodecki05}, Minkowski inequalities \cite{carlen99,carlen08}, or holographic gravity theory \cite{headrick07,hirata07}. There has also been recent interest in the structure of states where there is equality in the DPI \cite{hayden04,herbut04,jencova10}. Our approach provides a new perspective by decomposing the proof of the DPI into a simple proof of a more fundamental property, followed by a specialization. It also provides a new approach to teaching the DPI.

Most precisely, we first prove the DPI for a different entropy: the smooth min-entropy (Theorem~\ref{thm:subaddsme}). This proof is almost trivial and only involves the partial trace applied to the definition of the smooth min-entropy \cite{koenig08}. Then we can specialize the smooth min-entropy to the von Neumann entropy by the quantum asymptotic equipartition property (QAEP) (Theorem~\ref{thm:QAEP}) \cite{tomamichel08}. Here we provide a short proof that omits the analysis of the rate of convergence of this specialization, as apposed to \cite{tomamichel08}. We therefore obtain a self-contained proof for the von Neumann entropy DPI (Theorem~\ref{thm:subaddvN}).

We begin by introducing the smooth min-entropy (Section~2). This is followed by a high level proof of the data processing inequality for the von Neumann entropy (Section~3). Section~4 provides a proof of the QAEP. Finally Section~5 contains lemmas needed for the proofs in the previous sections.

%%%%%%%%%%%%%%%%%
\section{Smooth Min-Entropy}
%%%%%%%%%%%%%%%%%
\noindent
It has become apparent in recent works \cite{koenig08,tomamichel08,datta08,renner05} that smooth min-entropy is a relevant quantity for measuring quantum information. It characterizes operational tasks in information processing such as data compression and physics in the general one-shot setting, such as in statistical mechanics. Note that the one-shot setting does not make assumptions about the structure of relevant states, for example that they have product form. Since the von Neumann entropy also has an operational significance under certain additional assumptions, it could be expected that the von Neumann entropy can be obtained from smooth entropies as a special case. This is indeed true: the von Neumann entropy can be seen as an ``averaged" smooth entropy via the QAEP. We introduce a particular entropy, the min-entropy\footnote{It is sufficient to take the maximum over $\lambda$ if a finite dimensional system is considered. However, in infinite dimensions it is necessary to take a supremum \cite{furrer10}.}
\begin{equation}
H_{\min}(A|B)_{\rho}:=\max_{\lambda}\{\lambda\in\mathbb{R} \mid \exists \; \sigma_{B}\in S_{=}(\mathcal{H}_{B}) \text{ s.t. } \rho_{AB} \leq 2^{-\lambda} \mathbbm{1}_{A} \otimes \sigma_{B}\},
\end{equation}
which leads to the smooth min-entropy, defined as
\begin{equation}
H_{\min}^{\epsilon}(A|B)_{\rho} := \max_{\rho_{AB}' \in \mathcal{B}^{\epsilon}(\rho_{AB})} H_{\min}(A|B)_{\rho'}. \label{eq:Hmin1}
\end{equation}
The state $\sigma_{B}$ is chosen from the set of normalized states $S_{=}(\mathcal{H}_{B})$ in the Hilbert space $\mathcal{H}_{B}$. The state $\rho'_{AB}$ is chosen from the set of subnormalized states in the Hilbert space $\mathcal{H}_{AB}$ that are also close to the state $\rho_{AB}$: $\mathcal{B}^{\epsilon}(\rho_{AB}) := \left\{ \rho'_{AB} | \rho'_{AB} \in S_{\leq}(\mathcal{H}_{AB}) , P(\rho_{AB},\rho'_{AB}) \leq \epsilon \right\}$. To specify this $\epsilon$-ball around a state $\rho$, we use the purified distance \cite{tomamichel09} $P(\rho,\sigma) := \sqrt{1-F(\rho,\sigma)^2}$ (where $F(\rho,\sigma):=\norm{\sqrt{\rho}\sqrt{\sigma}}_{1}$ and $\norm{\rho}_{1}:=\Tr \sqrt{\rho \rho^{\dag}}$).\footnote{If $\rho$ and $\sigma$ are not normalized, then the generalized fidelity is used: $\bar{F}(\rho,\sigma) := \norm{\sqrt{\rho \oplus (1-\Tr \rho)}\sqrt{\sigma \oplus (1-\Tr \sigma)}}_{1}$. If either $\rho$ or $\sigma$ is normalized, then the generalized fidelity reduces to the standard fidelity.}

%%%%%%%%%%%%%%%%%%%%
\section{Data processing inequality}
%%%%%%%%%%%%%%%%%%%%
\noindent
We are now ready to state our main result and provide a high-level proof. If the entropies of interest are interpreted operationally then Theorem~\ref{thm:subaddsme} below deals with data processing in the one-shot scenario: a local physical operation is performed on a tri-partite quantum system \emph{once}, and a statement is made about the information content of such a system. Theorem~\ref{thm:subaddvN} can be interpreted as an average scenario: a statement is made about the information content \emph{on average} after applying a local physical operation to a tri-partite quantum state.

It is important to note that our proof of the DPI for the smooth min-entropy (Theorem \ref{thm:subaddsme}, below) applies to infinite- and finite-dimensional systems (see \cite{furrer10}), while our proof of the DPI for the von Neumann entropy (Theorem \ref{thm:subaddvN}, below) only applies to finite dimensions.

%%%%%%%%%%%%%%%%%%%%%%
\subsection{General Data Processing Inequality}
%%%%%%%%%%%%%%%%%%%%%%
\begin{theorem} [\cite{renner05,tomamichel09,koenig08} Smooth min-entropy DPI] \label{thm:subaddsme} Let $\rho \in S_{=}(\mathcal{H}_{ABC})$. Then
\begin{equation} \label{eq:dpi}
H_{\min}^{\epsilon}(A|BC)_{\rho} \leq H_{\min}^{\epsilon}(A|B)_{\rho}.
\end{equation}
\end{theorem}
\noindent
\proof{First we let $\lambda := H_{\min}^{\epsilon}(A|BC)_{\rho}$ and we choose the particular $\tilde{\rho}_{ABC} \in \ball{\rho_{ABC}}$ and $\sigma_{BC}$ in the definition of $H_{\min}^{\epsilon}(A|BC)_{\rho}$ such that $\lambda$ is maximized. From Eq.~\ref{eq:Hmin1} we have $\tilde{\rho}_{ABC} \leq 2^{-\lambda} \mathbbm{1}_{A} \otimes \sigma_{BC}$, and by tracing out system $C$, which is a positive map, we get $\tilde{\rho}_{AB} \leq 2^{-\lambda} \mathbbm{1}_{A} \otimes \sigma_{B}$. We know that $\tilde{\rho}_{ABC} \in \ball{\rho_{ABC}}$, and therefore $P(\rho_{ABC},\tilde{\rho}_{ABC})\leq \epsilon$. Since the purified distance does not increase under the partial trace (see Lemma~\ref{lemma:ballcpm}), it follows that $P(\rho_{AB},\tilde{\rho}_{AB})\leq \epsilon$. Therefore we have $\tilde{\rho}_{AB} \in \ball{\rho_{AB}}$, and $\sigma_{B}\in S_{=}(\mathcal{H}_{B})$, which are candidates for maximizing $H_{\min}^{\epsilon}(A|B)_{\rho}$.}

%%%%%%%%%%%%%%%%%%%%
\subsection{Specialized Data Processing Inequality}
%%%%%%%%%%%%%%%%%%%%
\noindent
Now we have completed the proof of the DPI in the most general case, and the only remaining difficulty is to specialize Theorem~\ref{thm:subaddsme} to the DPI for the von Neumann entropy. This specialization is achieved by using the limit of many i.i.d.\ copies of a state, called the QAEP.
\begin{theorem}[\cite{tomamichel08} QAEP]\label{thm:QAEP} Let $\rho\in S_{=}(\mathcal{H}_{AB})$. Then 
\begin{equation}
\lim_{\epsilon\to 0} \lim_{n\to\infty}\frac{1}{n}H^{\epsilon}_{\min}(A^n|B^n)_{\rho^{\otimes n}} = H(A|B)_{\rho}.
\end{equation}
\end{theorem}
This directly reduces Theorem~\ref{thm:subaddsme} to the DPI for the von Neumann entropy.
\begin{theorem}[\cite{lieb73,lieb73b,simon79,nielsen05,petz86,ruskai07,horodecki06,horodecki05,carlen99,carlen08,headrick07,hirata07} von Neumann entropy DPI]\label{thm:subaddvN} Let $\rho \in S_{=}(\mathcal{H}_{ABC})$. Then 
\begin{equation}
H(A|BC)_{\rho} \leq H(A|B)_{\rho}.
\end{equation}
\end{theorem}
However, in order to have a self contained proof of the data processing inequality for the von Neumann entropy we provide an alternative, shorter proof of the QAEP than that of \cite{tomamichel08}.

%%%%%%%%%%%%%%%%%%%%%%%
\section{Quantum Asymptotic Equipartition Property}
%%%%%%%%%%%%%%%%%%%%%%%
\noindent
In order to prove Theorem~\ref{thm:QAEP}, we upper and lower bound $\lim_{\epsilon\to 0}\lim_{n\to\infty}H^{\epsilon}_{\min}(A^n|B^n)_{\rho^{\otimes n}}$ by $H(A|B)_{\rho}$. These bounds rely on basic properties of smooth entropies, which will be proved in Section~5. The lower bound (Lemma~\ref{lemma:lower}) is obtained by applying a chain rule to the conditional smooth min-entropy such that it is bounded by a difference of non-conditional smooth entropies (Lemma~\ref{lemma:chain}). The i.i.d.\ limit of non-conditional smooth entropies can then be taken (Lemmas~\ref{lemma:lowerboundHminlimit} and~\ref{lemma:upperboundH0limit}). The upper bound (Lemma~\ref{lemma:upper}) can be obtained by bounding the smooth min entropy by the von Neumann entropy of a nearby state (Lemma~\ref{lemma:upperbound}), and then using the continuity of the von Neumann entropy when the i.i.d.\ limit is taken (Lemma~\ref{lemma:limitH}).

For these proofs we will need the smooth $0^{\text{th}}$ order R\'enyi entropy, which is defined as $H_{0}^{\epsilon}(A)_{\rho} := \min_{\rho' \in \mathcal{B}^{\epsilon}(\rho)} H_{0}(A)_{\rho'}$, where $H_{0}(A)_{\rho} := \log \rank \rho_{A}$. In addition, we will need the non-conditional smooth min-entropy defined as $H^{\epsilon}_{\min}(A)_{\rho}:=\max_{\rho'\in\ball{\rho}} H_{\min}(A)_{\rho'}$, where $H_{\min}(A)_{\rho} := -\log \norm{\rho_{A}}_{\infty}$. The infinity norm is defined as $\norm{\rho}_{\infty} := \max_{i} \{ | \lambda_{i} | \}$, where $\lambda_{i}$ are the eigenvalues of $\rho$. In addition, note that $H^{\epsilon}_{\min}(A|B)_{\rho}$ reduces to $H_{\min}(A)$ in the case that $B$ is trivial and $\epsilon=0$.

\begin{lemma}[Lower bound on the conditional smooth min-entropy]\label{lemma:lower}
Let $\rho\in S_{=}(\mathcal{H}_{AB})$. Then
\begin{equation}
H(A|B)_{\rho} \leq \lim_{\epsilon\to 0}\lim_{n\to\infty} \frac{1}{n} H^{\epsilon}_{\min}(A^{n}|B^{n})_{\rho^{\otimes n}}.
\end{equation}
\end{lemma}
\proof{We use the chain rule Lemma~\ref{lemma:chain} applied to the state $\rho\in S_{=}(\mathcal{H}_{AB})$:
\begin{equation}\label{eq:chainrule}
H_{\min}^{\frac{\epsilon}{3}}(AB)_{\rho} - H_{0}^{\frac{\epsilon}{3}}(B)_{\rho}  \leq H_{\min}^{\epsilon}(A|B)_{\rho}.
\end{equation}
Next we use the non-conditional QAEP of Lemmas~\ref{lemma:lowerboundHminlimit} and \ref{lemma:upperboundH0limit} given by
\begin{equation}\label{eq:qaep2}
H(A)_{\rho} \leq \lim_{\epsilon \to 0} \lim_{n \to \infty} \frac{1}{n}H_{\min}^{\epsilon}(A^n)_{\rho^{\otimes n}}, \quad H(A)_{\rho} \geq \lim_{\epsilon \to 0} \lim_{n \to \infty} \frac{1}{n}H_{0}^{\epsilon}(A^n)_{\rho^{\otimes n}}.
\end{equation}
We can apply Eq.~\ref{eq:chainrule} to the state $\rho^{\otimes n}$, divide by $n$, take the limit as $\epsilon\to 0$ and $n\to\infty$, and then use Eq.~\ref{eq:qaep2} to show that the left hand side is bounded by
\begin{equation}
H(A|B)_{\rho} \leq\lim_{\epsilon\to 0}\lim_{n\to\infty}\frac{1}{n}(H_{\min}^{\frac{\epsilon}{3}}( A^n B^n )_{\rho^{\otimes n}} - H_{0}^{\frac{\epsilon}{3}}(B^n )_{\rho^{\otimes n}}),
\end{equation}
where we use the definition of the conditional von Neumann entropy.}

\begin{lemma}[Upper bound on the conditional smooth min-entropy]\label{lemma:upper}
Let $\rho\in S_{=}(\mathcal{H}_{AB})$. Then
\begin{equation}
\lim_{\epsilon\to 0}\lim_{n\to\infty} \frac{1}{n} H^{\epsilon}_{\min}(A^{n}|B^{n})_{\rho^{\otimes n}} \leq H(A|B)_{\rho}.
\end{equation}
\end{lemma}
\proof{We apply the relation of conditional von Neumann entropy and conditional smooth min-entropy, Lemma~\ref{lemma:upperbound}, to the state $\rho^{\otimes n}_{A^{n}B^{n}}$:
\begin{equation}
H_{\min}^{\epsilon}(A^n|B^n)_{\rho^{\otimes n}} \leq H(A^n|B^n)_{\tilde{\rho}},
\end{equation}
where $\tilde{\rho}\in\ball{\rho^{\otimes n}_{AB}}$. Dividing by $n$, then taking the limit as $\epsilon\to 0$ and $n\to\infty$, and using the limit of the conditional von Neumann entropy of an almost i.i.d. state, Lemma~\ref{lemma:limitH}, we have:
\begin{equation}
\lim_{\epsilon\to 0} \lim_{n\to\infty}\frac{1}{n}H(A^n|B^n)_{\tilde{\rho}} = H(A|B)_{\rho}.
\end{equation}}

%%%%%%%%%%%%%%%%%%%%
\section{General Properties of Smooth Entropies}
%%%%%%%%%%%%%%%%%%%%
\noindent
The following are properties of smooth entropies used to prove Lemmas~\ref{lemma:lower}, and \ref{lemma:upper}. In particular, we bound the smooth min-entropy and smooth $0^{th}$-order R\'enyi entropy in order to perform the i.i.d.\ limit of $\epsilon\to 0$, $n\to \infty$. The proofs rely on certain basic properties of the von Neumann entropy and distance measures, which are provided in the appendices.

\begin{lemma}[Chain rule]\label{lemma:chain}
Let $\rho \in S_{=}(\mathcal{H}_{AB})$. Then 
\begin{equation}
H_{\min}^{\epsilon}(AB)_{\rho} - H_{0}^{\epsilon}(B)_{\rho} \leq H_{\min}^{3 \epsilon}(A|B)_{\rho}.
\end{equation}
\end{lemma}
\noindent
\proof{We pick the particular $\rho'_{AB} \in \ball{\rho_{AB}}$ in the definition of the non-conditional smooth min-entropy $H_{\min}^{\epsilon}(AB)_{\rho}=\lambda$ such that it is maximized.  We also pick the particular $\tilde{\rho}_{B}\in\ball{\rho_{B}}$ from the definition of the $0^{\text{th}}$ order R\'enyi entropy such that it is minimized, and write the projector onto its support as $\Pi:=\Pi_{\supp(\tilde{\rho}_{B})}$. Now given that $\rho'_{AB} \leq 2^{-\lambda} \mathbbm{1}_{AB}$, then $\Pi\rho'_{AB}\Pi \leq 2^{-\lambda} \mathbbm{1}_{A} \otimes \mathbbm{1}_{\supp(\tilde{\rho}_{B})}$, so we have
\begin{equation}
H^{\epsilon}_{\min}(AB)_{\rho}= \lambda, \quad \Pi \rho'_{AB} \Pi\leq 2^{-\lambda} \mathbbm{1}_{A} \otimes \mathbbm{1}_{\supp(\tilde{\rho}_{B})}.  \label{eq:minentrop} 
\end{equation}
\noindent
Now we will need to ensure that $\hat{\rho}_{AB}:=\Pi\rho'_{AB} \Pi$ is close to $\rho_{AB}$. To do this, we use the triangle inequality for the purified distance (see Lemma 5 of \cite{tomamichel09}) in the first and third lines, as well as the fact that the purified distance decreases under the CP trace non-increasing map \hbox{$\rho\to\Pi \rho \Pi$} (Lemma~\ref{lemma:ballcpm}) in the second line:
\begin{align}
P(\hat{\rho}_{AB},\rho_{AB}) &\leq P(\hat{\rho}_{AB},\tilde{\rho}_{AB}) + P(\tilde{\rho}_{AB},\rho_{AB}) \\
&\leq P(\rho'_{AB},\tilde{\rho}_{AB}) + P(\tilde{\rho}_{AB},\rho_{AB}) \\
&\leq P(\rho'_{AB},\rho_{AB}) + 2 P(\tilde{\rho}_{AB},\rho_{AB}) \\
&= \epsilon + 2 P(\tilde{\rho}_{AB},\rho_{AB}), \label{eq:halfwaythere}
\end{align}
where we purify $\tilde{\rho}_{B}$ to the state $\ket{\phi}_{ABC}$ and define $\tilde{\rho}_{AB} := \Tr_{C}(\ket{\phi}\bra{\phi})$ (see Lemma 8 of \cite{tomamichel09}). Now all that is left to find is $P(\tilde{\rho}_{AB},\rho_{AB})$. From Theorem~\ref{thm:uhlmann} we can define a purification $\ket{\psi}_{ABC}$ of $\rho_{B}$ such that $\Tr_{C} \ket{\psi}\bra{\psi}=\rho_{AB}$ and the following holds:
\begin{equation}
P(\ket{\phi}_{ABC},\ket{\psi}_{ABC}) = P(\tilde{\rho}_{B},\rho_{B}). \label{eq:purifyequiv}
\end{equation}
Now since the purified distance doesn't increase under the partial trace (see Lemma~\ref{lemma:ballcpm}):
\begin{equation}
P(\ket{\phi}_{ABC},\ket{\psi}_{ABC}) \geq P(\tilde{\rho}_{AB},\rho_{AB}) \geq P(\tilde{\rho}_{B}, \rho_{B}). \label{eq:ptraceineq}
\end{equation}
Combining Eqs.~\ref{eq:purifyequiv} and~\ref{eq:ptraceineq} we get
\begin{equation} \label{eq:purifnochange}
P(\ket{\phi}_{ABC},\ket{\psi}_{ABC}) = P(\tilde{\rho}_{AB},\rho_{AB}) = P(\tilde{\rho}_{B}, \rho_{B}).
\end{equation}
We know that $P(\tilde{\rho}_{B},\rho_{B})\leq \epsilon$, and therefore $P(\tilde{\rho}_{AB},\rho_{AB}) \leq \epsilon$. This makes Eq.~\ref{eq:halfwaythere} $P(\hat{\rho}_{AB},\rho_{AB}) \leq 3 \epsilon$. Now returning to the the smooth min-entropy in Eq.~\ref{eq:minentrop}, we define $\tau_{\tilde{\rho}_{B}} := \mathbbm{1}_{\supp (\tilde{\rho}_{B})} / \rank(\tilde{\rho}_{B})$ so that we have
\begin{align}
H^{\epsilon}_{\min}(AB)_{\rho}&= \left\{  \lambda+\log(\rank(\tilde{\rho}_{B})) \mid \Pi\rho'_{AB} \Pi\leq 2^{-\lambda} \mathbbm{1}_{A} \otimes \tau_{\tilde{\rho}_{B}} \right\} \\
&\leq \max_{\hat{\rho} \in \mathcal{B}^{3\epsilon}(\rho)} \max_{\sigma_{B}} \left\{ \lambda \mid \hat{\rho}_{AB} \leq 2^{-\lambda} \mathbbm{1}_{A} \otimes \sigma_{B} \right\} + \log(\rank(\tilde{\rho}_{B})) \\
&= H_{\min}^{3\epsilon}(A|B)_{\rho_{AB}} + H^{\epsilon}_{0}(B)_{\rho_{B}}.
\end{align}}

Now we provide some bounds on non-conditional smooth R\'enyi entropies by non-conditional R\'enyi entropy (Lemmas \ref{lemma:lowerboundHmin} and \ref{lemma:upperboundH0}). We then use these bounds to show one direction of the non-conditional QAEP (Lemmas \ref{lemma:lowerboundHminlimit} and \ref{lemma:upperboundH0limit}). Note that the non-conditional QAEP is known, and is sometimes referred to as Schumacher compression \cite{schumacher95}. It can be proved by using projectors onto a typical set. It can also be essentially reduced to a classical problem that can be shown using the law of large numbers \cite{cover91}. We provide our proofs below since they provide an alternative proof using bounds on smooth entropies in terms of R\'enyi entropies, and these bounds may be of general interest in quantum information theory.

\begin{lemma}[Lower bound on the smooth min-entropy]\label{lemma:lowerboundHmin}
Let $\rho\in S_{=}(\mathcal{H}_{A})$, $\alpha>1$, and $\epsilon\in (0,1]$. Then
\begin{equation}
H_{\alpha}(A)_{\rho} + \frac{\log (1-\sqrt{1-\epsilon^2})}{\alpha-1} \leq H_{\min}^{\epsilon}(A)_{\rho}.
\label{eq:limitpart2}
\end{equation}
\end{lemma}
\proof{First, we let $\rho=\sum_{x} \lambda_{x} \ket{x}\bra{x}$. We construct a quantum state $\sigma$ whose eigenvectors are the same as those of $\rho$, and whose eigenvalues, $\nu_{x}$, are $\nu_{x} = \lambda_{x}$ if $x\in\mathcal{X}$ and $\nu_{x} = 0$ otherwise, where $\mathcal{X}:=\{x \in \{ 1,2,\ldots,\dim \mathcal{H} \} : \lambda_{x} \leq \lambda^{*}\}$, and $\lambda^{*} \in \left[ 0 , 1 \right]$. Note that we will fix $\lambda^{*}$ to a specific value later in the proof. Hence $\sigma \in S_{\leq}(\mathcal{H})$. Now we may write the fidelity between $\rho$ and $\sigma$ as
\begin{equation}
\norm{\sqrt{\rho}\sqrt{\sigma}}_{1} = \sum_{x} \lambda_{x}^{1/2} \nu_{x}^{1/2} = \sum_{x \in \mathcal{X}} \lambda_{x}.
\end{equation}
We can write (for $\alpha>1$):
\begin{equation}
\sum_{x} \lambda_{x}^{\alpha} \geq \sum_{x\notin\mathcal{X}} \lambda_{x}^{\alpha-1}\lambda_{x} \geq \norm{\sigma}_{\infty}^{(\alpha-1)}\sum_{x\notin\mathcal{X}} \lambda_{x} = \norm{\sigma}_{\infty}^{(\alpha-1)}\left(1-F(\rho,\sigma)\right).
\end{equation}
By taking the $\log$ of this equation and since $\nu_{x}\leq\norm{\sigma}_{\infty}$ $\forall x$ we get
\begin{equation}
H_{\alpha}(A)_{\rho} \leq \frac{1}{1-\alpha} \log (1-F(\rho,\sigma)) + H_{\min}(A)_{\sigma}.
\end{equation}
Now we choose a particular $\lambda^{*}$ so that the fidelity is fixed to be $F(\rho,\sigma)=\sqrt{1-\epsilon^2}$ ($1\geq\epsilon>0$). This means that $P(\rho,\sigma) \leq \epsilon$, and hence $\sigma \in \mathcal{B}^{\epsilon}(\rho)$, so $H_{\min}(A)_{\sigma} \leq H^{\epsilon}_{\min} (A)_{\rho}$.}

\begin{lemma}[Upper bound on the $0^{th}$ order R\'enyi entropy]\label{lemma:upperboundH0}
Let $\rho\in S_{=}(\mathcal{H}_{A})$, $1/2<\alpha<1$, and $\epsilon\in [0,1)$. Then
\begin{equation}
H_{0}^{\epsilon}(A)_{\rho} \leq H_{\alpha}(A)_{\rho}+\frac{1}{\alpha-1}\log \sqrt{1-\epsilon}.
\end{equation}
\end{lemma}
\proof{This proof follows similarly to the proof of Lemma~\ref{lemma:lowerboundHmin}. We can construct a quantum state $\sigma$ in the same manner as Lemma~\ref{lemma:lowerboundHmin}. Now $1/2<\alpha<1$ so we have
$
\sum_{x} \lambda_{x}^{\alpha} \geq \sum_{x\in\mathcal{X}}\lambda_{x}^{\alpha} \geq (1/\rank\sigma)^{(\alpha-1)}\sum_{x\in\mathcal{X}} \lambda_{x}.
$
Taking the $\log$ gives
$
H_{\alpha}(A)_{\rho} \geq \frac{1}{1-\alpha} \log F(\rho,\sigma) + H_{0}(A)_{\sigma}.
$
Now we choose a particular $\lambda^{*}$ so that we can write the fidelity as $F(\rho,\sigma)=\sqrt{1-\epsilon}$, ($1>\epsilon\geq0$), and so $\sigma\in\ball{\rho}$. Then $H_{0}(A)_{\sigma} \geq H^{\epsilon}_{0} (A)_{\rho}$, which gives the result.}

\begin{lemma}[Non-conditional QAEP for smooth min-entropy]\label{lemma:lowerboundHminlimit}
Let $\rho\in S_{=}(\mathcal{H}_{A})$. Then
\begin{equation}
H(A)_{\rho} \leq \lim_{\epsilon \to 0} \lim_{n \to \infty} \frac{1}{n}H_{\min}^{\epsilon}(A^n)_{\rho^{\otimes n}}.
\end{equation}
\end{lemma}
\proof{First, we calculate the quantum R\'enyi entropy of order $\alpha$, defined as $H_{\alpha}(A)_{\rho} := 1/(1-\alpha) \log \Tr \rho^{\alpha}$ for the state $\rho^{\otimes n}$:
\begin{equation}
H_{\alpha}(A^n)_{\rho^{\otimes n}} = n H_{\alpha}(A)_{\rho}. \label{eq:tensoralpha}
\end{equation}
Now we may write Eq.~\ref{eq:limitpart2} from Lemma~\ref{lemma:lowerboundHmin} as
\begin{equation}
\lim_{\epsilon \to 0} \lim_{n \to \infty} \frac{1}{n}H_{\min}^{\epsilon}(A^n)_{\rho^{\otimes n}} \geq H_{\alpha}(A)_{\rho}.
\end{equation}
This is true for all $\alpha >1$ and so in particular, it's true if we take the limit as $\alpha\to 1^{+}$, where we know from Lemma~\ref{lemma:alphato1} that $\lim_{\alpha \to 1} H_{\alpha}(A)_{\rho} = H(A)_{\rho}$.}

\begin{lemma}[Non-conditional QAEP for $0^{th}$-order R\'enyi entropy]\label{lemma:upperboundH0limit}
Let $\rho\in S_{=}(\mathcal{H}_{A})$. Then
\begin{equation}
H(A)_{\rho} \geq \lim_{\epsilon \to 0} \lim_{n \to \infty} \frac{1}{n}H_{0}^{\epsilon}(A^n)_{\rho^{\otimes n}}.
\end{equation}
\end{lemma}
\proof{This follows in a similarly to the proof of Lemma~\ref{lemma:lowerboundHminlimit}, but now Lemma~\ref{lemma:upperboundH0} is used.}

\begin{lemma}[Relation of conditional von Neumann and conditional smooth min-entropy]\label{lemma:upperbound}
Let $\rho \in S_{=}(\mathcal{H}_{AB})$. Then $\exists \; \tilde{\rho} \in \ball{\rho}$ such that
\begin{equation}
H^{\epsilon}_{\min}(A|B)_{\rho} \leq H(A|B)_{\tilde{\rho}}.
\end{equation}
\end{lemma}
\proof{We start with the definition of the conditional von Neumann entropy for subnormalized states $\tilde{\rho}_{AB}\in S_{\leq}(\mathcal{H}_{AB})$, so we have
\begin{align}
H(A|B)_{\tilde{\rho}}&:=\frac{1}{\Tr \tilde{\rho}_{AB}} \max_{\sigma_{B}} \Tr (\tilde{\rho}_{AB}(\log(\mathbbm{1}_{A} \otimes \sigma_{B})-\log(\tilde{\rho}_{AB}))) \\
&\geq \frac{1}{\Tr \tilde{\rho}_{AB}} \Tr ( \tilde{\rho}_{AB}(\log(\lambda \mathbbm{1}_{A} \otimes \sigma'_{B}) -\log(\tilde{\rho}_{AB}))) - \log \lambda,
\end{align}
where we drop the maximization, picking a specific $\sigma_{B}'$: the state that allows $\lambda$ to be maximized in $H_{\min}(A|B)_{\rho}$. We have also added and subtracted $\log \lambda$, defined as $-\log \lambda=H_{\min}^{\epsilon}(A|B)_{\rho}$, and we choose $\tilde{\rho}$ to be the state that allows $\lambda$ to be maximized in the definition of $H_{\min}^{\epsilon}(A|B)_{\rho}$.
Also, to simplify our expression, we use the quantum relative entropy, defined as $H(\rho||\sigma) := \Tr (\rho \log \rho) - \Tr (\rho \log \sigma)$. Now we may write
\begin{equation}
-\frac{1}{\Tr \tilde{\rho}_{AB}} H(\tilde{\rho}_{AB} || \lambda \mathbbm{1}_{A} \otimes \sigma'_{B}) + H_{\min}^{\epsilon}(A|B)_{\rho}\geq H_{\min}^{\epsilon}(A|B)_{\rho},
\end{equation}
where in the last line, we use the monotonicity of the $\log$ to show that $\tilde{\rho}_{AB}\log \tilde{\rho}_{AB} \leq \tilde{\rho}_{AB} \log (\lambda \mathbbm{1}_{A} \otimes \sigma'_{B})$. This then implies $-H(\tilde{\rho}_{AB} || \lambda \mathbbm{1}_{A} \otimes \sigma'_{B}) \geq 0$.}

\nonumsection{Acknowledgements}
\noindent
The authors acknowledge support from the European Research Council (grant no. 258932), and the Swiss National Science Foundation (grant no. 200020-135048).

\nonumsection{References}

%\bibliographystyle{unsrt}
%\bibliography{DPIQIC}

\appendix{: Known Distance Properties}
\noindent
The following are known properties used in the proof of Theorem~\ref{thm:QAEP}, which we include here for completeness.
\noindent
\begin{theorem}[\cite{nielsen00,uhlmann76}  Uhlmann's Theorem] \label{thm:uhlmann}
Let $\rho,\sigma \in S_{=}(\mathcal{H})$. Then
\begin{equation}
F(\rho,\sigma)=\max_{\ket{\psi},\ket{\phi}} |\bra{\psi} \phi \rangle| = \max_{\ket{\phi}} |\bra{\psi} \phi \rangle|,
\end{equation}
where $\ket{\phi},\ket{\psi}$ are purifications of $\rho$ and $\sigma$ respectively.
\end{theorem}

\begin{lemma}[\cite{tomamichel09} Purified distance under CP trace non-increasing maps]\label{lemma:ballcpm}
Let $\mathcal{E}$ be a trace non-increasing map, and $\rho,\sigma \in S_{\leq}(\mathcal{H})$. Then
\begin{equation}
P(\mathcal{E}(\rho),\mathcal{E}(\sigma)) \leq P(\rho,\sigma).
\end{equation}
\end{lemma}
This can be proven by using the fact that the generalized fidelity cannot decrease under completely positive trace non-increasing maps.

\begin{lemma}[\cite{berta10} Purified distance relation]\label{lemma:eig} Let $\rho,\sigma\in S_{=}(\mathcal{H})$, and let $r_{i}$ and $s_{i}$ be their eigenvalues respectively in non-increasing order ($r_{i+1}\leq r_{i}$ and $s_{i+1}\leq s_{i}$ $\forall i$). Also, define $\tilde{\sigma}:=\sum_{i}s_{i}\ket{i}\bra{i}$, where $\ket{i}$ are the eigenvalues of $\rho$. Then
\begin{equation}
P(\rho,\sigma)\geq P(\rho,\tilde{\sigma})
\end{equation}
\end{lemma}
\proof{ Showing that $F(\rho,\sigma)\leq F(\rho,\tilde{\sigma})$ is sufficient, as the result then follows from the definition of the purified distance. From the definition of the fidelity we have
\begin{equation}
F(\rho,\sigma) = \max_{U}\text{Re} \Tr (U\sqrt{\rho}\sqrt{\sigma}) \leq \max_{U,V}\text{Re} \Tr (U\sqrt{\rho}V\sqrt{\sigma}) = \sum_{i}\sqrt{r_{i}}\sqrt{s_{i}} = F(\rho,\tilde{\sigma}),
\end{equation}
where the maximizations are taken over all unitaries, and Theorem~7.4.9 and Eq.~7.4.14 are used from \cite{horn85}.
}

\appendix{: Known Entropic Properties}
\noindent

\begin{lemma}[Limit of the conditional von Neumann entropy of an almost i.i.d.\ state]\label{lemma:limitH}
Let $\rho \in S_{=}(\mathcal{H}_{AB})$ and $\sigma_{n} \in \ball{\rho^{\otimes n}}$. Then
\begin{equation} \label{eq:limiteq2}
\lim_{\epsilon \to 0} \lim_{n \to\infty} \frac{1}{n}H(A^n|B^n)_{\sigma_{n}} = H(A|B)_{\rho}.
\end{equation}
\end{lemma}
\proof{First, we know that $\sigma_{n} \in \ball{\rho^{\otimes n}}$, and by Eq.~\ref{eq:purifnochange} we have $P(\rho^{\otimes n}_{B},\sigma_{n_{B}})\leq \epsilon$. Now we show Eq.~\ref{eq:limiteq2} is valid when the system $B$ is trivial, \ie $H(A^{n}|B^{n})_{\sigma_{n}}=H(A^{n})_{\sigma_{n}}$ and $H(A|B)_{\rho}=H(A)_{\rho}$ (see Chapter 3 of \cite{cover91}). 

We extend $\rho^{\otimes n}_{A}$ and $\sigma_{n_{A}}$ to $\rho'_{n} := \rho^{\otimes n}_{A} \oplus 0$ and $\sigma'_{n} := \sigma_{n_{A}} \oplus (1-\Tr \sigma_{n_{A}})$ so that $\sigma'_{n} \in S_{=}(\mathcal{H}_{A}\oplus \mathcal{H}_{1})$ (where $\mathcal{H}_{1}$ is a one dimensional space). Next, we define the state $\tilde{\sigma}_{n} := \sum_{i} s'_{i} \ket{i} \bra{i}$, where $s'_{i}$ are the eigenvalues of $\sigma'_{n}$ ordered such that $s'_{i} \geq s'_{i+1}, \forall i$ and $\ket{i}$ are the eigenvectors of $\rho'_{n}$. It is clear that $P(\rho^{\otimes n}_{A},\sigma_{n_{A}}) = P(\rho'_{n},\sigma'_{n})$, and so by Lemma~\ref{lemma:eig}, we know that $P(\rho'_{n},\sigma'_{n})\geq P(\rho'_{n},\tilde{\sigma}_{n})$. The purified distance is lower bounded by the trace distance \cite{fuchs99}, and so $P(\rho'_{n},\tilde{\sigma}_{n})\geq D(\rho'_{n},\tilde{\sigma}_{n})$. Now since $\sigma_{n_{A}} \in \ball{\rho^{\otimes n}_{A}}$ we know $D(\rho'_{n},\tilde{\sigma}_{n})\leq \epsilon$. Now we may use Fannes' Inequality \cite{fannes73}:
\begin{align}
&\lim_{\epsilon \to 0} \lim_{n \to\infty} \frac{1}{n} \left | H(A^n)_{\tilde{\sigma}_{n}} - H(A^n)_{\rho'_{n}} \right |  \label{eq:etasum}\\
&\leq \lim_{\epsilon \to 0} \lim_{n \to\infty} \frac{1}{n} ( \epsilon \log d^n + \eta (\epsilon) )  = 0,
\end{align}
where we define $\eta(x) := -x\log x$, and $d=\dim(\mathcal{H}_{A})$. This is not the limit we would like to know, so we compare the entropies here to those of Eq.~\ref{eq:limiteq2} for trivial $B$. From the definition of $\tilde{\sigma}_{n}$ we know that $H(A^n)_{\tilde{\sigma}_{n}} = H(A^n)_{\sigma_{n}} - \eta(1-\Tr \sigma_{n})$ and so
\begin{align}
&\lim_{\epsilon\to 0} \lim_{n\to \infty} \frac{1}{n} |H(A^n)_{\rho^{\otimes n}} - H(A^n)_{\sigma_{n}}| \label{eq:eqn80}\\
&\leq \lim_{\epsilon\to 0} \lim_{n\to \infty} \frac{1}{n} (|H(A^n)_{\rho'_{n}} \!-\! H(A^n)_{\tilde{\sigma}_{n}}| \!+\! |\eta(1-\Tr \sigma_{n})|) \nonumber\\
&=0, \nonumber
\end{align}
where we know that $0\leq(1-\Tr\sigma_{n_{A}})\leq 1$, and hence $0\leq \eta(1-\Tr\sigma_{n_{A}})\leq 1/2$.

When $B$ is non-trivial we can combine Eq.~\ref{eq:eqn80} with Eq.~\ref{eq:tensoralpha} and the definition of the conditional von Neumann entropy to get the result.}

\begin{lemma}[\cite{cover91} Relation of R\'enyi entropy and von Neumann entropy]\label{lemma:alphato1}
Let $\rho \in S_{=}(\mathcal{H}_{A})$. Then
\begin{equation}
\lim_{\alpha \to 1} H_{\alpha}(A)_{\rho} = H(A)_{\rho}.
\end{equation}
\end{lemma}

\end{document}